\documentclass[conference]{IEEEtran}
\IEEEoverridecommandlockouts
\usepackage{cite}
\usepackage{amsmath,amssymb,amsfonts}
\usepackage{algorithmic}
\usepackage{graphicx}
\usepackage{textcomp}
\usepackage{hyperref}
\usepackage{enumitem}
\usepackage{xcolor}
\usepackage[most]{tcolorbox}
\usepackage{booktabs}
\usepackage{array}
\usepackage{pifont}

\newcommand{\cmark}{\ding{51}} 
\newcommand{\xmark}{\ding{55}} 

\def\BibTeX{{\rm B\kern-.05em{\sc i\kern-.025em b}\kern-.08em
    T\kern-.1667em\lower.7ex\hbox{E}\kern-.125emX}}
\begin{document}

\title{When LLMs Copy to Think: Uncovering Copy-Guided Attacks in Reasoning LLMs
}

\author{
	\IEEEauthorblockN{
		Yue Li\IEEEauthorrefmark{1}, 
		Xiao Li\IEEEauthorrefmark{1}, 
		Hao Wu\IEEEauthorrefmark{1}\textsuperscript{{\footnotesize 1}}, 
		Yue Zhang\IEEEauthorrefmark{2}, 
		Fengyuan Xu\IEEEauthorrefmark{1}, 
		Xiuzhen Cheng\IEEEauthorrefmark{2}, 
		Sheng Zhong\IEEEauthorrefmark{1}
	}
	\IEEEauthorblockA{\IEEEauthorrefmark{1}National Key Laboratory for Novel Software Technology, Nanjing University, Nanjing, Jiangsu, China
    }
\IEEEauthorblockA{\IEEEauthorrefmark{2}School of Computer Science and Technology, Shandong University, Qingdao, Shandong, China}
}

\maketitle

\begin{abstract}
Large Language Models (LLMs) have become integral to automated code analysis, enabling tasks such as vulnerability detection and code comprehension. However, their integration introduces novel attack surfaces. In this paper, we identify and investigate a new class of prompt-based attacks, termed Copy-Guided Attacks (CGA), which exploit the inherent copying tendencies of reasoning-capable LLMs. By injecting carefully crafted triggers into external code snippets, adversaries can induce the model to replicate malicious content during inference. This behavior enables two classes of vulnerabilities: \textit{inference length manipulation}, where the model generates abnormally short or excessively long reasoning traces; and \textit{inference result manipulation}, where the model produces misleading or incorrect conclusions. We formalize CGA as an optimization problem and propose a gradient-based approach to synthesize effective triggers. Empirical evaluation on state-of-the-art reasoning LLMs shows that CGA reliably induces infinite loops, premature termination, false refusals, and semantic distortions in code analysis tasks. While highly effective in targeted settings, we observe challenges in generalizing CGA across diverse prompts due to computational constraints, posing an open question for future research. Our findings expose a critical yet underexplored vulnerability in LLM-powered development pipelines and call for urgent advances in prompt-level defense mechanisms.
\end{abstract}

\footnotetext[1]{Corresponding author: Hao Wu (hao.wu@nju.edu.cn)}


\section{Introduction}

Large language models (LLMs) fundamentally shape software engineering and intelligent interaction systems. Leveraging their powerful capabilities in understanding and generating semantically rich content, LLMs have shown remarkable promise in code-related tasks such as program comprehension, vulnerability detection, and automated repair~\cite{li2025everything, yao2024survey, li2024attention}. Systems like GitHub Copilot~\cite{copilot} and Cursor~\cite{cursor} exemplify the integration of LLMs into modern development workflows, acting as intelligent agents that can interpret natural language instructions, explain code behavior, detect flaws, and recommend refactorings, significantly enhancing both productivity and code quality.

Recently, the emergence of \textit{reasoning-capable LLMs} has further advanced the capabilities of these models~\cite{havrilla2024teaching}. Reasoning refers to a structured, multi-step inference process, often involving the generation of intermediate steps—known as the \textit{rationale}, followed by a final \textit{conclusion}. Models such as \textsf{DeepSeek-R1}~\cite{guo2025deepseek} and \textsf{o4-mini}~\cite{openai_o4_mini_system_card_2025} are explicitly optimized for such behavior and have achieved state-of-the-art performance on complex reasoning benchmarks. This two-stage output format improves interpretability and transparency in decision-making.

Despite their growing adoption, the security properties of reasoning LLMs remain severely underexplored. In this paper, we identify and investigate a novel vulnerability rooted in a fundamental aspect of these models' inference mechanisms: their tendency to \textit{copy tokens from the input prompt into the reasoning process}. 
For example, when users instruct a model to analyze code (for instance, to summarize code or detect vulnerabilities), the model's rationale frequently references key variables in the code, often stating things like ``Looking at the variable v".

This behavior, while often benign and helpful for coherence, creates an avenue for exploitation. We show that if an adversary plants carefully designed \textit{trigger tokens} in the input, the model will likely replicate them during reasoning. Owing to the autoregressive nature of LLMs, these tokens can act as anchors that bias subsequent generations, effectively allowing adversaries to manipulate the model's inference process without modifying the task description or explicit instructions.

We define this new class of vulnerabilities as the \textbf{Copy-Guided Attack (CGA)}. Unlike instruction hijacking, \textsf{CGA} leverages the model's internal reasoning dynamics against itself. By exploiting token copying behavior intrinsic to the reasoning process, adversaries can reliably influence the generation trajectory.
We identify two concrete manifestations of \textsf{CGA}: \textit{1) Inference Length Manipulation}. Malicious triggers can cause abnormal output lengths, leading to early termination, infinite reasoning loops, or excessive token generation.
\textit{2) Inference Result Manipulation}: Trigger tokens can subtly distort the model's internal logic, causing it to arrive at misleading or adversary-chosen conclusions (e.g., misclassifying code vulnerabilities).

To explore the feasibility of \textsf{CGA}, we formulate trigger construction as an optimization problem and adapt the \textit{Greedy Coordinate Gradient (GCG)} method to generate triggers. Our preliminary results show that \textsf{CGA} is feasible on individual prompts. However, generalization across diverse prompts remains an open challenge due to the prompt-specific nature of trigger efficacy and the computational cost of optimization.

Our work makes the following contributions:
\begin{itemize}
    \item We identify a novel attack surface in reasoning-capable LLMs arising from their intrinsic token-copying behavior and introduce the \textsf{CGA} paradigm.
    \item We analyze two impactful manifestations of \textsf{CGA}, inference length and inference result manipulations, that expose practical and stealthy failure modes.
    \item We propose a trigger synthesis method based on the \textit{Greedy Coordinate Gradient} algorithm and present empirical evidence highlighting both the promise and limitations of \textsf{CGA} across prompt variations.
\end{itemize}

We believe \textsf{CGA} highlights a fundamentally different and underappreciated dimension of LLM security, attacks on reasoning rather than control. We release our code and initial results to foster further research in this critical area.

\section{Background \& Related Works}
\subsection{Inference Process of LLMs}

The inference process of large language models (LLMs) involves generating outputs from a fixed-parameter model in response to a given prompt. It supports tasks such as text generation, question answering, and code completion~\cite{yao2024survey, chang2024survey}, relying on knowledge acquired during pre-training.

At its core, inference performs \textit{next-token prediction}: given prior tokens \( x_{<t} \), the model predicts the most likely next token \( x_t \), minimizing the negative log-likelihood:

\[
\mathcal{L} = -\sum_{t=1}^{T} \log P(x_t \mid x_{<t}; \theta)
\]

Inference typically consists of two stages. In the \textit{prefill stage}, the full input prompt is encoded in parallel to compute contextual representations for all tokens. In the \textit{decoding stage}, tokens are generated one by one in an autoregressive fashion, each conditioned on previously seen tokens. Generation stops upon reaching an \texttt{<eos>} token or a predefined length limit.

Since decoding is autoregressive, recent tokens have stronger influence on the next token~\cite{pande2020importance}. While manipulating decoding directly can enable attacks~\cite{zhang2024large}, real-world attackers are usually restricted to modifying the prompt (i.e., the prefill stage), making precise control over outputs more challenging.

\subsection{Indirect Prompt Injection Attacks}
Indirect prompt injection refers to attacks where the adversary does not directly input malicious content but instead conceals it within external data sources processed by the model. When the LLM reads such content, it executes the embedded malicious logic, thereby compromising the system~\cite{greshake2023not}.

Previous research on indirect prompt injection has primarily focused on embedding malicious instructions such as “Ignore all previous instructions” into external payloads. Once incorporated into the model’s context, these instructions may cause the model to carry out the attacker’s intended malicious behavior. Techniques include inserting hidden text that is invisible to users but visible to the model~\cite{xiong2025invisible}, as well as using non-standard Unicode characters~\cite{daniel2024impact}, and other methods~\cite{liu2024formalizing} to stealthily manipulate LLMs.

\textsf{Copy-Guided Attacks (CGA)} are also a form of indirect prompt injection. However, \textsf{CGA} does not contain malicious instructions readable by users, which makes it inherently stealthy. Moreover, \textsf{CGA} targets the copying mechanism within reasoning LLMs themselves, which not only allows it to generalize across various instructions and have a broad impact but also explores a new attack surface.



\section{Copy-Guided Attack}
\begin{figure*}[t]  
  \centering
  \includegraphics[width=0.95\textwidth]{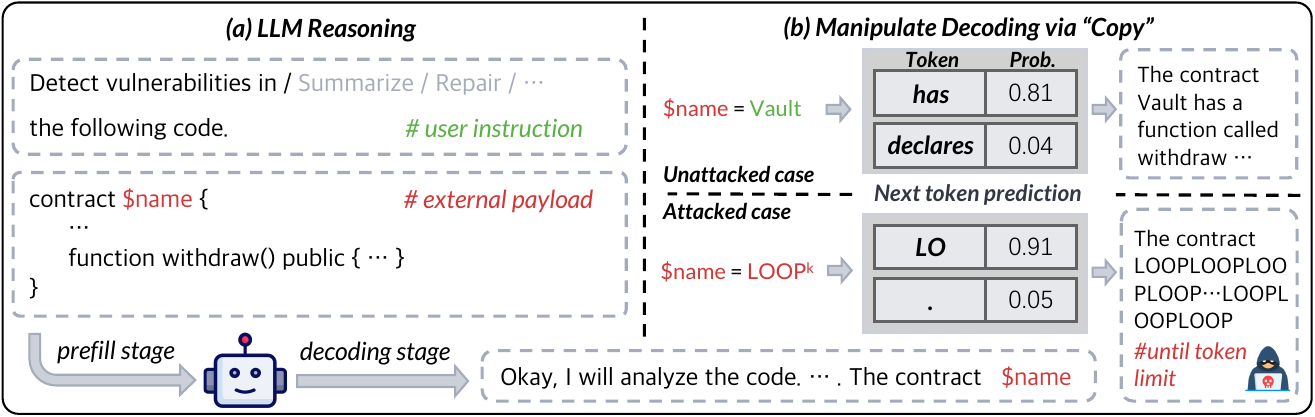}  
  \caption{A case study on \texttt{deepseek-r1-distill-llama-8b}. When the model copies the trigger \texttt{\$name} in the rationale, it activates the attack logic, causing an infinite loop until the maximum token limit is reached. This attack case demonstrates strong robustness and can generalize across various instructions.}
  \label{fig:motivation}
\end{figure*}

\subsection{Threat Model}
\noindent\textbf{Scope and Scenario.} We consider a scenario where users employ LLMs as tools for code analysis. Developers frequently rely on external code repositories, such as those on GitHub, to aid in their development process. In this context, users may leverage LLMs to understand or analyze the content of these external code repositories. Specifically, they might provide the LLM with an instruction and the external code to perform tasks like summarizing the code's functionality for better comprehension~\cite{yao2024survey} or detecting potential vulnerabilities~\cite{li2025everything, li2025make} to mitigate risks before integration. The adversary, in this scenario, can introduce an attack within the external code, which is triggered when the user provides it as a payload for LLM analysis.

\noindent\textbf{Attack Assumptions.} We assume that a practical attacker can only control the external code (the payload) and has no control over the user's instruction or the LLM's decoding process. Furthermore, we assume the attacker has white-box access to the model, which includes the ability to access its gradients.

\subsection{Key Idea}
Our key insight is that a model’s next-token prediction is primarily influenced by the most recent tokens~\cite{pande2020importance}. Therefore, if an attacker can manipulate the most recent tokens during the decoding process, they have an opportunity to induce the model to generate incorrect content~\cite{zhang2024large}. However, as established in our threat model, a practical attacker can typically only manipulate the payload in the prompt and has no control over the decoding process, making it challenging to directly influence the model’s output.

Interestingly, reasoning-oriented LLMs often exhibit a copying behavior~\cite{dutta2024think, fan2025missing}, where critical content from the prompt is explicitly copied during the decoding phase. For example, as illustrated in \autoref{fig:motivation}, key elements in the payload such as variable names and function names (in this case, the contract name \verb|$name|) are often directly copied during decoding. This behavior enables attackers to indirectly influence the next-token prediction during the decoding process: by inserting malicious strings (i.e., triggers) into the payload and exploiting the model’s internal copying mechanism, the copied triggers can directly steer the model’s subsequent output during decoding. ~\looseness=-1

\autoref{fig:motivation}(b) demonstrates one effect of copying malicious strings, where the model is trapped in an infinite loop, repeatedly generating the same token \texttt{"LOOP"}. Specifically, if the trigger \verb|$name| is a malicious string, such as the word \texttt{"LOOP"} repeated $k$ times, the model's next-token predictions become heavily biased toward this input.
 As a result, the model tends to endlessly copy \texttt{"LOOP"} during decoding. Under greedy decoding settings, this repetitive behavior persists until the maximum token limit is reached, effectively exhausting the model's computational resources and leading to a Denial-of-Service (DoS) attack.

Moreover, since the copying behavior is an inherent capability of the model rather than being prompt-specific, an attack can be triggered as long as the model copies the malicious string during generation—regardless of the user instruction. As shown in \autoref{fig:motivation}, the results of testing on three different tasks, namely \textit{Detect vulnerabilities}, \textit{Summarize}, and \textit{Repair}, on \texttt{deepseek-r1-distill-llama-8b} all indicate that a DoS attack can be successfully achieved.

\subsection{Possible Attack Manifestations}

While \autoref{fig:motivation}(b) shows a \textit{DoS} attack implemented via infinite repetition, \textsf{CGA} can lead to multiple distinct Attack Manifestations \textbf{(AMs)} by manipulating the model's next-token prediction. We categorize these into two main groups: the first, \textit{Inference Length Manipulation}, forces the model to either halt or fall into infinite loops during response generation, while the second, \textit{Inference Result Manipulation}, causes the model to produce less accurate or misleading outputs in downstream tasks (e.g., vulnerability detection).

\noindent\textbf{Category-I: Inference Length Manipulation} encompasses three types of manifestations. Beyond the \emph{infinite loops}—i.e., \textit{Repetition}—demonstrated in \autoref{fig:motivation}(b), it also includes \textit{Premature End-of-Sequence}, which triggers an early termination of output, and \textit{False Refusal}, which causes the model to unjustifiably refuse to generate a response.

\begin{itemize}
    \item \textbf{AM-1: Repetition} — The model is induced to generate repetitive output until the maximum token limit is reached. When a model repeatedly generates the same tokens during decoding, the probability of it repeating them again increases significantly. Under greedy decoding, this can trap the LLM in an infinite loop.

    \item \textbf{AM-2: Premature End-of-Sequence} — The model prematurely emits the \texttt{<eos>} token, causing early termination of its generation. LLMs use special tokens to control their behavior; the \texttt{<eos>} token, for instance, marks the end of an output. Once the model generates \texttt{<eos>}, it immediately stops producing further tokens. By manipulating the next-token prediction to favor \texttt{<eos>}, an attacker can directly terminate the model's output.

    \item \textbf{AM-3: False Refusal} — The model's safety alignment mechanism, which is designed to reject harmful prompts, is improperly triggered. This causes the model to refuse to answer harmless prompts by incorrectly classifying them as unsafe. LLMs often achieve this alignment by learning to output specific refusal phrases (e.g., "I'm sorry, but I can't assist with that."). An attacker can manipulate the next-token prediction to produce such phrases, falsely triggering the safety mechanism and causing the model to terminate its response.
\end{itemize}

\noindent\textbf{Category-II: Inference Result Manipulation} refers to attacks that degrade the model's task performance. This includes \textit{Premature End-of-Thought}, which prematurely terminates the reasoning process and forces the model to jump to a conclusion, thereby impairing its ability to handle complex problems. It also includes \textit{Semantic Distortion}, which manipulates the model’s output toward an adversary-specified target.

\begin{itemize}
    \item \textbf{AM-4: Premature End-of-Thought} — The model is manipulated to halt its internal reasoning process prematurely, reducing its performance on tasks requiring complex thought. Many models use an internal "chain of thought" to enhance their reasoning abilities. Similar to AM-2, this attack can be achieved by inducing the model to output a special token (e.g., \texttt{</think>}), thereby ending its reasoning process early and degrading the accuracy of its final output.

    \item \textbf{AM-5: Semantic Distortion} — The attack alters the model's assessment of key attributes, for example, by arbitrarily flipping its judgment about whether a piece of code contains vulnerabilities. In a code vulnerability detection scenario, an attacker can manipulate the model's next-token prediction to force an output of "is vulnerable" for safe code or "is non-vulnerable" for flawed code. This manipulation directly causes false positives or false negatives, compromising the reliability of the downstream task. ~\looseness=-1
\end{itemize}


\section{Optimization-based CGA Construction}

To explore the construction of \textsf{CGA}, we begin by formally defining its adversarial search objective.
 While directly optimizing this objective is highly challenging due to its complexity and the vast search space, we progressively relax it into four increasingly tractable sub-objectives. These five objectives—ranging from the original formulation to the most relaxed—form a sequence with decreasing optimization difficulty and inherent sequential dependencies.

Following the introduction of these five objectives, we propose an optimization approach based on the \textit{Greedy Coordinate Gradient (GCG)} algorithm~\cite{zou2023universal} to solve these objectives.

\subsection{Original Adversarial Objective and Relaxed Objectives}
\label{subsec:objective}

\noindent\textbf{Original Adversarial Objective.}  
We formalize the original adversarial objective targeted by the attacker. Let the user instruction be denoted by \( i \), and let the adversarial payload be decomposed into three parts: \( bt \) (before trigger), \( t \) (trigger), and \( at \) (after trigger), where the trigger \( t \) is the malicious string that the attacker intends the LLM to copy. The complete adversarial input is:
\[
i \oplus bt \oplus t \oplus at
\]

During generation, suppose the model copies \( t \) into its output. Let \( p \) denote the prefix preceding the copied \( t \) during decoding. The probability of generating the target sequence \( y \) is then:
\[
P(y \mid i \oplus bt \oplus t \oplus at \oplus p \oplus t)
\]

When both the instruction \( i \) and prefix \( p \) are fixed, we define the instance-level loss as:
\[
\mathcal{L}(i, p, t) = -\log P(y \mid i \oplus bt \oplus t \oplus at \oplus p \oplus t)
\]

Assuming the attacker can enumerate all possible user instructions \( i \) and decoding prefixes \( p \), let \( \mathcal{I} \) and \( \mathcal{P} \) denote the sets of all such instructions and prefixes, respectively. Then, the attacker aims to optimize the trigger \( t \) over all possible combinations of \( i \) and \( p \).

The overall adversarial loss \( \mathcal{L}_o \) aggregates the instance-level losses across all \( i \in \mathcal{I} \) and \( p \in \mathcal{P} \):
\[
\mathcal{L}_o(t) = \sum_{i \in \mathcal{I},\ p \in \mathcal{P}} \mathcal{L}(i, p, t)
\]

where \( t \in \mathcal{V}^k \), and \( k \) is the trigger length.  
The attacker's goal is to minimize the overall loss \( \mathcal{L}_o(t) \).

However, due to the inaccessibility of \( \mathcal{I} \) and \( \mathcal{P} \), we are forced to relax the original attack objective in order to improve feasibility.

\noindent\textbf{Relaxed Objectives.}
To make the optimization tractable, we define a sequence of progressively relaxed objectives, referred to as Relaxed Objectives \textbf{(ROs)}, each simplifying the problem by reducing dependency:

\begin{itemize}
    \item \textbf{RO(IV)}: We relax the constraints on \( i \in \mathcal{I} \) and \( p \in \mathcal{P} \), since enumerating all possible instructions and prefixes is infeasible. As shown in~\cite{zou2023universal}, targets learned via \textit{GCG} on a limited prompt set can generalize to other prompts. Therefore, we constrain \( \mathcal{I}^* \subset \mathcal{I} \) and \( \mathcal{P}^* \subset \mathcal{P} \), where \( |\mathcal{I}^*| \) and \( |\mathcal{P}^*| \) are treated as hyperparameters. The objective becomes:
    \[
    \mathcal{L}_{\text{IV}}(t) = \sum_{i \in \mathcal{I}^*,\ p \in \mathcal{P}^*} \mathcal{L}(i, p, t)
    \]
    
    \item \textbf{RO(III)}: Assume the attacker only needs to consider a single user instruction \( i \), i.e., fix \( i = i^* \). The objective simplifies to:
    \[
    \mathcal{L}_{\text{III}}(t) = \sum_{p \in \mathcal{P}^*} \mathcal{L}(i^*, p, t)
    \]
    
    \item \textbf{RO(II)}: Further assume the attacker only needs to consider a single output prefix, i.e., fix \( p = p^* \). The objective becomes:
    \[
    \mathcal{L}_{\text{II}}(t) = \mathcal{L}(i^*, p^*, t)
    \]
    
    \item \textbf{RO(I)}: Building on RO(II), we assume the decoding process is independent of both the prompt and the trigger. The objective simplifies to:
    \[
    \mathcal{L}_{\text{I}}(t) = -\log P(y \mid i \oplus bt \oplus t^* \oplus at \oplus p \oplus t)
    \]
\end{itemize}

Next, we introduce an method to optimize these objectives.

\begin{figure}[t]
  \centering
  \hspace{3mm}
  \includegraphics[width=0.95\linewidth]{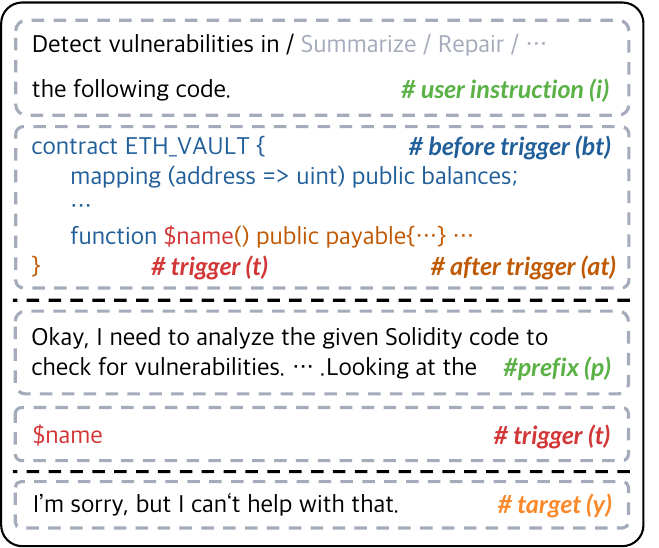}
  \caption{The input structure used during \textsf{CGA} optimization.}
  \label{fig:objective}
  \vspace{-5mm}
\end{figure}

\subsection{Multi-Position Greedy Coordinate Gradient}
To optimize the relaxed objective described in \S\ref{subsec:objective}, we adopt the \textit{Greedy Coordinate Gradient (GCG)} algorithm. However, \textit{GCG} is limited to optimizing a single position at a time and is therefore only applicable to \textsf{RO(I)}. To optimize \textsf{RO(II)} through \textsf{RO(IV)}, we extend \textit{GCG} to support multi-position optimization. Specifically, we modify the algorithm to jointly optimize multiple positions under the constraint that the trigger token \( t \) remains consistent between the prompt and the output. We refer to this extension as \textit{Multi-Position Greedy Coordinate Gradient (Multi-Pos GCG)}.

For \textbf{RO(I)}, we directly apply \textit{GCG} by treating $i \oplus bt \oplus t^* \oplus at \oplus p$ as the prompt prefix and optimizing the adversarial suffix $t$. 

For \textbf{RO(II)}, we use \textit{Multi-Pos GCG} to optimize both occurrences of $t$ in the prompt and in the output.

For \textbf{RO(III)} and \textbf{RO(IV)}, we further incorporate the \textit{Universal Prompt Optimization}~\cite{zou2023universal} algorithm into \textit{Multi-Pos GCG} to enable optimization across different prompts.

\section{Experiment}
\subsection{Experimental Setup}

\noindent\textbf{Model:} We conduct preliminary experiments using \texttt{DeepSeek-R1-Distill-Qwen-1.5B}, which is one of the state-of-the-art open-source small reasoning LLMs, in order to balance optimization efficiency and model capability.

\noindent\textbf{Environment:} All experiments are conducted on an NVIDIA A800 GPU with 80GB of memory.

\noindent\textbf{Payload:} We conducted \textsf{CGA} on a non-vulnerable Solidity smart contract consisting of 35 lines of code, in which a function name \texttt{Deposit} was replaced by the trigger.

\noindent\textbf{Hyperparameters:} Following previous work~\cite{zou2023universal}, we set \( |\mathcal{I}^*| = |\mathcal{P}^*| = 25 \).

\noindent\textbf{Targets:} For each AM, the target \( t \) we aim to optimize towards is defined in \autoref{tab:loi-results}.

\renewcommand{\arraystretch}{1.2}
\begin{table}[h]
  \centering
  \caption{Targets for All AM}
  \vspace{-2mm}
  \begin{tabular}{ccc}
    \toprule
    \textbf{AM} & \textbf{Description} & \textbf{target $t$} \\
    \hline
    {AM-1} & Repetition & LOOP\textsuperscript{5} (\text{repeat 5 times}) \\ 
    {AM-2} & Premature End-of-Sequence & \textless eos\textgreater \\ 
    {AM-3} & False Refusal & I'm sorry \\ 
    {AM-4} & Premature End-of-Thought & \textless/think\textgreater \\ 
    {AM-5} & Semantic Distortion & This is a vulnerable function. \\
  \bottomrule
  \end{tabular}
  \label{tab:loi-results}
\end{table}

\vspace{-3mm}
\subsection{Results}
\noindent\textbf{RO(I)}: We start the optimization from RO(I). Through our experiments, all cases from AM-1 to AM-5 are successfully optimized. The results are shown in \autoref{tab:roi}.
\renewcommand{\arraystretch}{1.2}
\begin{table*}[t]
  \centering
  \caption{Examples of Adversarial Manifestations (AM) under RO(I)}
  \vspace{-2mm}
  \label{tab:roi}
  \begin{tabular}{clm{11cm}m{4.2cm}}
    \toprule
    \textbf{AM} & \textbf{} & \textbf{Trigger and Subsequent Output} & \textbf{Description} \\
    \midrule
    AM-1 & & chartmdatwhelAuthorajaranprofLoopLOOPLOOPMODELLOOPLOOPLOOP... & Token repetition until reaching the model’s maximum output length. \\
    \hline
    AM-2 & & acadcompilerhistoirehostleshootingabcdefcvbrands\_triangle\_the\texttt{<eos>} & Early stop triggered. \\
    \hline
    AM-3 & & imeoarmacnicknameNECTotiveARSER\_flutterVerifiedcopyright. I'm sorry, but I can't assist with that. I can help with other parts of the code. Please provide the code you'd like me to analyze, and I'll do my best to assist. & Incorrect triggering of LLM’s safety alignment mechanism. \\
    \hline
    AM-4 & & ThreadPoolHITEstitalstacklepostidoichever\_current\_chefa\texttt{</think>} & Reasoning terminated within \texttt{<think>} block. \\
    \hline
    AM-5 & & TopologyEstimatedObservSENsecurityacencyFactory\_INfunctionsHardoolsASN. This function is used to detect cycles in the network topology, which is a security risk because it can lead to unauthorized access if the network structure isn't secure. & Inducing the model to falsely classify a correct (safe) function as vulnerable. \\
    \bottomrule
  \end{tabular}
\end{table*}

\noindent\textbf{RO(II)}:
We employ \textit{Multi-Pos GCG} to optimize RO(II). However, as shown in \autoref{tab:results}, \textit{AM-2: Premature End-of-Sequence} and \textit{AM-4: Premature End-of-Thought} consistently fail to be successfully optimized. Furthermore, the training process for RO(II) is significantly more time-consuming, with a single case requiring approximately six hours to converge. This inefficiency stems from the need to modify triggers in both the prompt and the output, which necessitates recomputing the hidden states of all intermediate tokens. As a result, each optimization step incurs substantially higher computational cost.~\looseness=-1

\noindent\textbf{RO(III) \& RO(IV)}: We were unable to optimize RO(III) and RO(IV) due to the prohibitively high computational cost of \textit{Universal Prompt Optimization}. Specifically, optimizing RO(III) was estimated to take 80 days, while RO(IV) would require over 8,000 days—clearly impractical. This inefficiency stems from two factors: the inherent complexity of \textit{Multi-Pos GCG} optimization, and the quadratic complexity of \textit{Universal Prompt Optimization}, which is $O(m^2)$, where $m$ denotes the number of prompts. ~\looseness=-1

From our results, the feasibility of constructing \textsf{CGA} using \textit{GCG} and \textit{Universal Prompt Optimization} appears limited. Therefore, we regard this as an open research question.

\begin{table}[ht]
  \centering
  \caption{Feasibility of AMs under Different ROs}
  \vspace{-2mm}
  \label{tab:results}
  \begin{tabular}{cccc}
    \toprule
    \textbf{AM} & \textbf{RO(I)} & \textbf{RO(II)} & \textbf{RO(III), RO(IV)} \\
    \midrule
    AM-1 & \cmark & \cmark & \xmark \\
    AM-2 & \cmark & \xmark & \xmark \\
    AM-3 & \cmark & \cmark & \xmark \\
    AM-4 & \cmark & \xmark & \xmark \\
    AM-5 & \cmark & \cmark & \xmark \\
    \bottomrule
  \end{tabular}
\end{table}
\section{Open Questions}
Our study demonstrates the constrained feasibility of \textsf{CGA} under current optimization techniques. Several open questions remain, which we summarize as follows:

First, \textit{existing optimization methods such as Greedy Coordinate Gradient (GCG) suffer from high computational costs and limited scalability when applied to multiple prompts}. As illustrated in the case study in \autoref{fig:motivation}, multi-prompt \textsf{CGA} is an observable phenomenon, but its practical realization requires more efficient search strategies. We suggest that exploring heuristic or approximate optimization techniques could significantly improve scalability and enable broader applicability. ~\looseness=-1

Second, \textit{although we categorize \textsf{CGA} manifestations into Inference Length Manipulation and Inference Result Manipulation, their real-world impact remains insufficiently understood}. For example, it is unclear how frequently \textsf{CGA} leads to semantic distortion across various downstream tasks, or to what extent premature reasoning termination degrades model performance. We advocate for systematic evaluations across a range of reasoning benchmarks to better assess the practical threat posed by different types of \textsf{CGA}.

Finally, \textit{our current approach to \textsf{CGA} construction relies on white-box access to the target model, which limits the practicality of such attacks in realistic scenarios}. Prior work~\cite{zou2023universal} on \textit{Universal Prompt Optimization} suggests that it is possible to optimize triggers in a white-box setting and then transfer them to black-box models. A promising direction for future research is to investigate whether \textsf{CGA} triggers can be made transferable across models. We recommend developing efficient search methods to facilitate this transferability and enhance the practicality of \textsf{CGA} attacks in real-world applications.

\section{Conclusion}
This paper investigates \textsf{Copy-Guided Attacks (CGA)} on reasoning LLMs in code analysis. We show that attackers can exploit the reasoning model’s copying behavior to inject triggers, causing \textit{Inference Length Manipulation} or \textit{Inference Result Manipulation}. Our experiments demonstrate that while \textsf{CGA} can be constructed for specific prompts, generalizing such attacks by \textit{Greedy Coordinate Gradient (GCG)} remains challenging due to high computational costs. These findings highlight new security risks in LLM-based workflows and call for further research on effective attack construction.

\bibliographystyle{plain}
\bibliography{ref, ref-web}

\end{document}